\documentstyle[12pt]{article}    % Specifies the document style.
\begin{document}           % End of preamble and beginning of text.
\rm
\baselineskip=0.33333in
\begin{quote} \raggedleft TAUP 2513-98
\end{quote}
\vglue 0.4in
\begin{center}{\bf The Role of Inert Objects\\
in Quantum Mechanical Phase}

\end{center}
\begin{center}E. Comay 
\end{center}
 
\begin{center}
School of Physics and Astronomy \\
Raymond and Beverly Sackler Faculty \\
of Exact Sciences \\
Tel Aviv University \\
Tel Aviv 69978 \\
Israel
\end{center}
\vglue 0.5in
\noindent
Email: eli@tauphy.tau.ac.il
\vglue 0.5in
\noindent
PACS No: 03.65.Bz

\vglue 0.5in
\noindent
Abstract:

   Quantum mechanical foundations of the polarized neutron
phase shift experiment are discussed. The fact that the neutron
retains its ground state throughout the experiment is shown to be
crucial for the phase shift obtained.

\newpage

   Experimental data of the interaction of the neutron's
magnetic moment with a magnetized material have two aspects:
they provide information on the structure of specific materials
and help us understand physical properties of the neutron. Thus,
neutron scattering off a magnetized material 
was used for deciding
that the neutron's magnetic moment is analogous to that of a
current loop[1-3]. A new aspect of this issue is discussed
in this work.

   Very sensitive results of neutron dynamics are derived from an
analysis of its interference pattern.
Consider the nonrelativistic electromagnetic
Lagrangian[4] augmented by
the interaction of the neutron's intrinsic magnetic moment
{\boldmath $\mu$}
with the external magnetic field {\boldmath $B$} [5]
\begin{equation}
L = \frac{1}{2}mv^2 + e \mbox{\boldmath $v\cdot A$} - eV +
\mbox{\boldmath $\mu \cdot B$}. 
\label{eq:LAGQ}
\end{equation}
In this work expressions are written in units where $c = 1$. 
Obviously, in the case of a neutron, the electric charge $e$ vanishes
and we are left with the first and the last terms of 
$(\!\!~\ref{eq:LAGQ})$. Hence, the action[4] 
derived from $(\!\!~\ref{eq:LAGQ})$ depends on the
external magnetic field, and one obtains
\begin{equation}
dS = (\frac{1}{2}mv^2 + \mbox{\boldmath $\mu \cdot B$}) dt.
\label{eq:DS}
\end{equation}
This expression shows how the magnetic field affects the action, and
thereby the neutron's phase and its interference pattern.
Recent experiments utilize this relation between interference
pattern and the Lagrangian $(\!\!~\ref{eq:LAGQ})$[6,7] for
understanding the dynamics of the system. The present work
analyses results of [7] and shows why quantum mechanical
properties of the neutron distinguish it from an ordinary
classical current loop.

   In the experiment reported in [7],
polarized neutrons drift
through an external time dependent magnetic field. The neutron's velocity
and spin as well as the external magnetic field are parallel to 
each other. At the neutron's location, the magnetic field
is practically uniform. Hence, neither force nor torque are exerted
on the neutron. The experiment confirms that,
in spite of these facts and due to
the action $(\!\!~\ref{eq:DS})$,
the neutron is affected
by the magnetic field which induces a phase shift that modifies
the neutron's interference pattern.

   Evidently, the neutron behaves in the experiment mentioned
above like an inert object. Indeed, as a quantum mechanical system,
the neutron is not excited to higher baryonic states whose energy
is several hundreds Mev above the ground state, by its very small
interaction with the external magnetic field. It is shown in this
work that this property of
the neutron is crucial for the magnetically dependent
phase shift of $(\!\!~\ref{eq:DS})$.

   Let us examine a hypothetical experiment which is similar to the
one described in [7]. All parts of this experiment are like those
of [7], except the neutron which is replaced by a "classical neutron"
whose structure is described below.

   Consider a ring made of an insulating material. This material
takes the shape of a thin circular pipe containing a positively charge
fluid. The insulating material is charged uniformly with negative
charge so that the device looks 
like an electrically neutral object. 
The charge to mass ratio of the fluid is very small. This fluid
rotates frictionlessly in a clockwise direction. Let $a$ denote the
ring's radius and $I$ is the electric current
associated with the fluid's motion. The magnetic field of
the device is like that of a tiny magnetic dipole[8]
\begin{equation}
\mbox{\boldmath $\mu $} = \pi a^2I\mbox{\boldmath $k$}
\label{eq:MURING}
\end{equation}
where {\boldmath $k$} denotes a unit vector in the $z$-direction.
This hypothetical experiment is carried out in conditions
where the nonrelativistic limit holds. The rather small ring 
is a macroscopic object and the motion of the charged fluid can be
treated classically. 

The system's Lagrangian is the ordinary Lagrangian of an electromagnetic
system[4], namely $(\!\!~\ref{eq:LAGQ})$ without its last term
\begin{equation}
L = \frac{1}{2}mv^2 + e \mbox{\boldmath $v\cdot A$} - eV.
\label{eq:LAGC}
\end{equation}
Here the electric potential $V$ vanishes. Hence, for evaluating
$(\!\!~\ref{eq:LAGC})$ we have to find
the values of the mechanical part and that of the second one which
is associated with the magnetic interaction.
The mechanical part of the Lagrangian describing the present
experiment consists of two terms. One term pertains to the motion
of the device in the $z$-direction and the second one 
is associated with
the rotation of the fluid. Hence, the required Lagrangian is
\begin{equation}
L = \frac{1}{2}Mv_z^2 + \frac{1}{2}m_fv_\perp ^2 +
                \int \mbox{\boldmath $j\cdot A$}d^3r
\label{eq:LAGC1}
\end{equation}
where $M$ and $m_f$ denote the mass of the entire device
and of the rotating fluid, respectively and
$v_\perp$ is the fluid's velocity in the $(x,y)$ plane.
The last term of $(\!\!~\ref{eq:LAGC1})$ is the continuum
analog of the single particle expression $e${\boldmath $v\cdot A$}
(see [8], p. 596).

   Let us evaluate the last term of $(\!\!~\ref{eq:LAGC1})$.
Introducing the
electric current $I$, one applies Stokes theorem, the
spatial uniformity of the magnetic field {\boldmath $B$} 
and relation $(\!\!~\ref{eq:MURING})$
and finds
\begin{eqnarray}
\int \mbox{\boldmath $j\cdot A$}d^3r & = & I\oint_L 
\mbox{\boldmath $A\cdot $}d\mbox{\boldmath $l$}
\nonumber \\
& = & I\int_S (curl
\mbox{\boldmath $A)\cdot $}d\mbox{\boldmath $s$}
\nonumber \\
& = & \pi a^2 IB
\nonumber \\
& = & \mbox{\boldmath $\mu\cdot B$}.
\label{eq:JA}
\end{eqnarray}
Here the integral subscript
$L$ denotes the closed path along the ring and $S$
is the ring's area.
It follows that in $(\!\!~\ref{eq:LAGC1})$,
the term {\boldmath $j\cdot A$} 
of the classical device discussed here
is analogous to the last term of
$(\!\!~\ref{eq:LAGQ})$.

   Unlike the neutron case, where its internal structure does
not change while the external magnetic field is turned on, the
kinetic energy of the rotating fluid depends on the external
magnetic field. Since the charge to mass ratio of the rotating
fluid is very small, the current $I$ is regarded in
the following calculation as a constant.
The power transmitted to the rotating charged fluid is
\begin{eqnarray}
P  & = & I\oint_L \mbox{\boldmath $E\cdot $}d\mbox{\boldmath $l$}
\nonumber \\
& = & I\int_S (curl \mbox{\boldmath $E)\cdot $}d\mbox{\boldmath $s$}
\nonumber \\
& = & -I\pi a^2 \frac {\partial B}{\partial t}
\label{eq:POWER}
\end{eqnarray}
where the spatial uniformity of {\boldmath $B$} is used.
Applying the notation of $(\!\!~\ref{eq:LAGC1})$,
let $v_\perp (T_0)$ denote the velocity of the charged fluid at $T_0$
before the external magnetic field is turned on. 
It follows that, at an instant $t$, this part of
the kinetic energy of the fluid is
\begin{eqnarray}
\frac {1}{2} mv_\perp ^2(t)  & = & \frac {1}{2} mv_\perp ^2(T_0) - I\pi a^2 
\int \frac {\partial B}{\partial t} dt
\nonumber \\
& = & \frac {1}{2} mv_\perp ^2(T_0) - I\pi a^2B.
\nonumber \\
& = & \frac {1}{2} mv_\perp ^2(T_0) - \mbox{\boldmath $\mu\cdot B$}.
\label{eq:DK}
\end{eqnarray}
Substituting $(\!\!~\ref{eq:JA})$ and $(\!\!~\ref{eq:DK})$
into $(\!\!~\ref{eq:LAGC1})$, one finds that in the classical
analog of the neutron experiment, 
the value of the Lagrangian 
\begin{equation}
L(t) = \frac{1}{2}Mv_z^2 + \frac{1}{2}m_fv_\perp ^2(T_0) =L(T_0)
\label{eq:LAGCT}
\end{equation}
is independent of the external magnetic field. The same is true
for the corresponding action.

   The results of this work emphasize the quantum mechanical
meaning of the polarized neutron interference experiment[7].
As stated in [7], the neutron is free of classical effects like
force and torque. In spite of this, its quantum properties vary
due to the external magnetic field. The discussion carried out
above points out another quantum mechanical aspect of the experiment.
The neutron's spin and its associated magnetic moment are
properties of a quantum mechanical system whose state may
vary only in quantum leaps. The excited baryonic states of the
neutron are very high, so that in experiment [7], the neutron
is always in its ground state and
behaves as an inert object. As such, there exists no analog to the
variation of the self energy of the rotating fluid
$(\!\!~\ref{eq:DK})$. For this reason, the Lagrangian of the
neutron experiment [7] and the corresponding action {\em depend}
on the external magnetic field whereas in the analogous
classical experiment discussed above they are {\em independent}
of it. It can be concluded that a comparison of the two experiments,
namely the actual experiment reported in [7] and the
hypothetical one which uses a classical current loop,
demonstrates another quantum mechanical aspect of the neutron
experiment [7].

%????
\newpage
References:
\begin{itemize}
\item[{[1]}] C. G. Shull, E. O. Wollan and W. A. Strauser,
Phys. Rev. {\bf 81} (1951) 483.
\item[{[2]}] F. Mezei, Physica {\bf 137B-C} (1986) 295.
\item[{[2]}] F. Mezei, Physica {\bf 151B-C} (1988) 74.
\item[{[4]}] L. D. Landau and E. M. Lifshitz, {\em The Classical
Theory of Fields} (Pergamon, Oxford, 1975). p. 46.
\item[{[5]}] J. M. Blatt and V. F. Weisslopf,
{\em Theoretical Nuclear Physics} (Wiley, New York, 1952) p. 32.
\item[{[6]}] J. Summhammer et al.,
Phys. Rev. Lett. {\bf 75} (1995) 3206.
\item[{[7]}] W. -T Lee, O. Motrunich, B. E. Allman and S. A. Werner,
Phys. Rev. Lett. {\bf 80} (1998) 3165.
\item[{[8]}]  J. D. Jackson, {\em Classical Electrodynamics,}
second edition (John Wiley, New York, 1975). pp. 178, 179.

\end{itemize}

\end{document}